# Modulation of Charge Transport and Rectification Behavior in CsSnI$_3$ Thin Films Through A-site Cation Engineering

Anna A. Zarudnyaya[1], Gleb V. Segal[1], Andrey P. Morozov[1], Lev O. Luchnikov[1], Sergey Yu. Yurchuk[2], Ivan V. Schemerov[2], Pavel A. Gostishchev[1*] and Danila S. Saranin[1*]

[1]LASE – Laboratory of Advanced Solar Energy, NUST MISIS, 119049 Moscow, Russia

[2]Department of semiconductor electronics and device physics, NUST MISIS, 119049 Moscow, Russia

**Corresponding authors:**

Dr. Pavel A. Gostishchev gostischev.pa@misis.ru,

Dr. Danila S. Saranin saranin.ds@misis.ru

**Abstract**

CsSnI$_3$ perovskite is a promising thin-film semiconductor with high intrinsic conductivity for various device applications (thermoelectric, photovoltaics, etc.). Stoichiometric CsSnI$_3$ owns high-density of defects and structural imperfections affecting device performance. In this work, we made an investigation on A-site cation engineering to evaluate the correlation between structural and transport parameters for effective operation in rectifying devices. Here, we analyzed CsSnI$_3$ thin films modified with methylamine (MA), formamidine(FA), guanidine(GuA) and 5-ammonium valeric acid (AVA) cations, correlating structural parameters obtained by Rietveld refinement with their optoelectronic and diode characteristics. MA-, FA-, and GuA-substituted films exhibited low sheet resistance (~450–2200 Ohm/sq); however, strain-induced lattice distortions and accumulated defects in GuA-substituted films significantly hindered effective charge collection and increased recombination losses. AVA substitution formed low-conductivity 2D interlayers, dramatically increasing resistance (>10$^5$ Ohm/sq) and altering transient response characteristics, yet provided minimal reverse-switching losses (~100 μW/cm²), beneficial for high-frequency applications. FA substitution emerged as optimal, balancing structural stability, conductivity, minimal defects, and superior diode properties. The obtained results highlight that targeted lattice modifications strongly influence the practical performance of rectifying p-i-n diodes based on CsSnI$_3$.

Iodide perovskites are among the most promising materials for next-generation energy conversion technologies [1,2]. Hybrid semiconductors with the general formula $ABI_3$ (where A represents an organic cation such as $CH_3NH_3^+$, $CH(NH_2)_2^+$, or $Cs^+$; B is a divalent metal cation such as $Pb^{2+}$, $Sn^{2+}$, or $Ge^{2+}$) exhibits a unique combination of optoelectronic and charge transport properties in thin-film configurations with microcrystalline morphology[3]. Owing strong optical absorption in the visible spectrum (up to $10^5$ cm$^{-1}$) and charge diffusion lengths extending several microns, Pb-based perovskites were considered as a promising absorber for solar cells and photodiodes. Typical compositions for photovoltaic applications, such as $FAPbI_3$ has the bandgap ($E_g$) of 1.5 eV and average carrier density of $10^{16}$ cm$^{-3}$[4,5], which allows to reach high power conversion efficiencies (PCE)[6] and/or specific detectivity ($D^*$)[7] of the devices, competitive to ones based on monocrystalline silicon[8]. However, the toxicity of lead could be a critical obstacle[9]. In this context, Sn-based compositions could be a relevant alternative[10]. Inorganic perovskites, such as $CsSnI_3$, mitigates the challenges associated with lead toxicity, and represents competitive electric characteristics caomparable with epitaxially grown materials: charge carrier mobility up to $10^2$ cm$^2$V$^{-1}$s$^{-1}$ (p-type) and lifetime values in the nanosecond range[11]. However, the specific features of inorganic tin-based perovskites points to different favorable applications. As reported, inorganic tin-based perovskites have increased carrier density up to $10^{17}$ cm$^{-3}$, large Seebeck coefficient ~$10^2$ μV/K and exceptionally low lattice thermal conductivity ~0.4 W/m[12,13]. This put $CsSnI_3$ as a strong candidate for thermoelectric applications, particularly in thin-film energy conversion devices fabricated via solution-based deposition techniques.

The optoelectronic performance of $CsSnI_3$ critically depends on its structural phase, with distinct differences between the metastable perovskite polymorphs. The black perovskite phases of $CsSnI_3$ (Orthorhombic – Pnma; Cubic - Pm3̄m) demonstrate strong charge transport properties, while oxidized derivatives (Black quasi 2D Phase - $Cs_2SnI_6$, Yellow Phase – Y-γ-CsSnI3,) suffers from compromised conductivity and optical absorption[14,15]. Particularly, Sn-cation has rapid oxidation dynamics. Surface Sn atoms oxidize first, forming $Sn^{4+}$-rich layers that propagate into the bulk of thin-films[16]. Oxidation processes induce lattice contraction, which triggers polymorphic phase transitions. Moreover, Cs-cation with ionic radius 1.88 Å mismatches the cuboctahedral void size in $SnI_6$ frameworks (ideal radius = 1.76 Å) and produces the residual lattice strain that destabilizes the phase composition [17,18]. Current literature reports suggest that partial relaxation of lattice strain can be achieved through substitution with organic cations of varying ionic radii, such as methyl ammonium (MA$^+$), and larger spacer molecules like ThDMA (2,5-thiophenedimethylammonium)[19]. However, a comprehensive understanding of the effects of such modifications requires not only an analysis of the chemical composition and structural properties but also a detailed comparison of their impact on charge transport characteristics, which is missing of inorganic $CsSnI_3$.

In this work, we present a comprehensive study on the correlation between cationic engineering of $CsSnI_3$ and the transport characteristics of thin films. Various cationic modifiers, including guanidine, amino acids, and others, were incorporated into thin films. Structural and surface modifications were systematically analyzed and compared with lateral and bulk conductivity, capacitive behavior, and rectification properties in the p-i-n diode architectures.

To investigate the impact of the A-site cation engineering on $CsSnI_3$, we used methylamine (MA, $CH_3NH_3$), formamidine (FA, $CH(NH_2)_2$), guanidine (GuA, $CH_6N_3$) and 5-ammonium valeric acid cation (AVA, $HOOC-(CH_2)_4-NH_3$). Each organic cation influences the lattice differently. A-site cation engineering in $CsSnI_3$ leverages diverse cation sizes (steric effects) and functional groups (electronic/chemical modulation) to tailor lattice strain and charge transport. MA/FA with a close match to the ionic radii of Cs could optimize lattice strain and affect the bandgap. On the other hand, guanidinium and 5-ammonium valeric acid are substantially larger than Cs; at high concentrations, they cannot fit into the cubic cage without breaking the 3D framework of $CsSnI_3$.

We fabricated A-site modified $CsSnI_3$ thin-films using solution processing with spin-coating technique. Briefly, 20% of the Cs concentration in the perovskite was substituted with A-site modifiers (MA, FA, GuA, AVA). To simplify sample identification, we used the following titles of the samples, corresponding to the used modifier: $MA_{0.2}Cs_{0.8}SnI_3$, $FA_{0.2}Cs_{0.8}SnI_3$, $GuA_{0.2}Cs_{0.8}SnI_3$, $AVA_{0.2}Cs_{0.8}SnI_3$. The thickness of the fabricated samples was measured with stylus profilometry. The values were measured in the range of 490-610 nm, as presented in **tab.1.** To estimate the impact of cation engineering integration on the general optic properties in the fabricated samples, we measured the absorbance spectra and analyzed Tauc plots[20] (**Fig.1 a.**). The extracted data showed that the use of large AVA cation tends to increase of $E_g$ up to 1.38 eV, while use of small methyl amine reduced to 1.27 eV.

Table 1 – Thickness of the fabricated samples

| **Composition** | **$MA_{0.2}Cs_{0.8}SnI_3$** | **$AVA_{0.2}Cs_{0.8}SnI_3$** | **$FA_{0.2}Cs_{0.8}SnI_3$** | **$GuA_{0.2}Cs_{0.8}SnI_3$** |
|---|---|---|---|---|
| Thickness, nm | 557 | 566 | 492 | 615 |

**Fig.1b** presents the surface potential distributions of $OC_{0.2}Cs_{0.8}SnI_3$ (OC – organic cation) samples via Kelvin probe force microscopy (**KPFM**) mapping (**Fig. S1** in electronic supplementary material, **ESI**). The measured KPFM maps corresponded to the atomic force microscopy of the films (**Fig. S2, ESI**). The surface potential values ($V_{CPD}$) for the samples containing MA, FA, and GuA cations were measured as -586 mV, -513 mV, and -725 mV, respectively. In the bias voltage regime (BV = 1 V), the calculation of the work function ($W_f$) could be done using equation: $W_f^{sample} = W_f^{tip} - V_{CPD}^{sample}$, where $W_f^{tip}$ is a $W_f$ of a cantilever's tip. The decrease in $V_{CPD}$ corresponds to an increase in the absolute values of $W_f$. We observed a 471 mV rise in surface potential in AVA$_{0.2}$Cs$_{0.8}$SnI$_3$, consistent with Fermi level pinning[21,22]. For GuA$_{0.2}$Cs$_{0.8}$SnI$_3$ samples, we observed a drop to 725 mW, which highlights the specific difference in the surface properties for introduction of various organic cations.

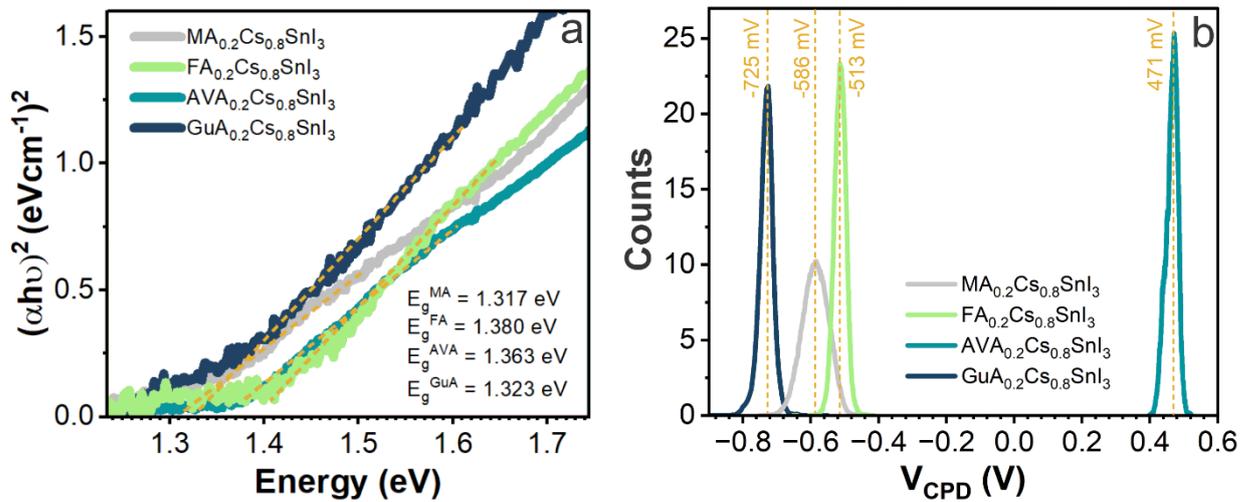

Figure 1 – Tauc plot (a) and surface potential distribution (b) of OC$_{0.2}$Cs$_{0.8}$SnI$_3$ samples

Next, we analyzed the changes of structural properties and lattice parameters using X-ray diffraction spectroscopy (**XRD**). **Fig. 2** presents the X-ray diffractograms of the modified CsSnI$_3$ thin-films deposited on glass with ITO coating.

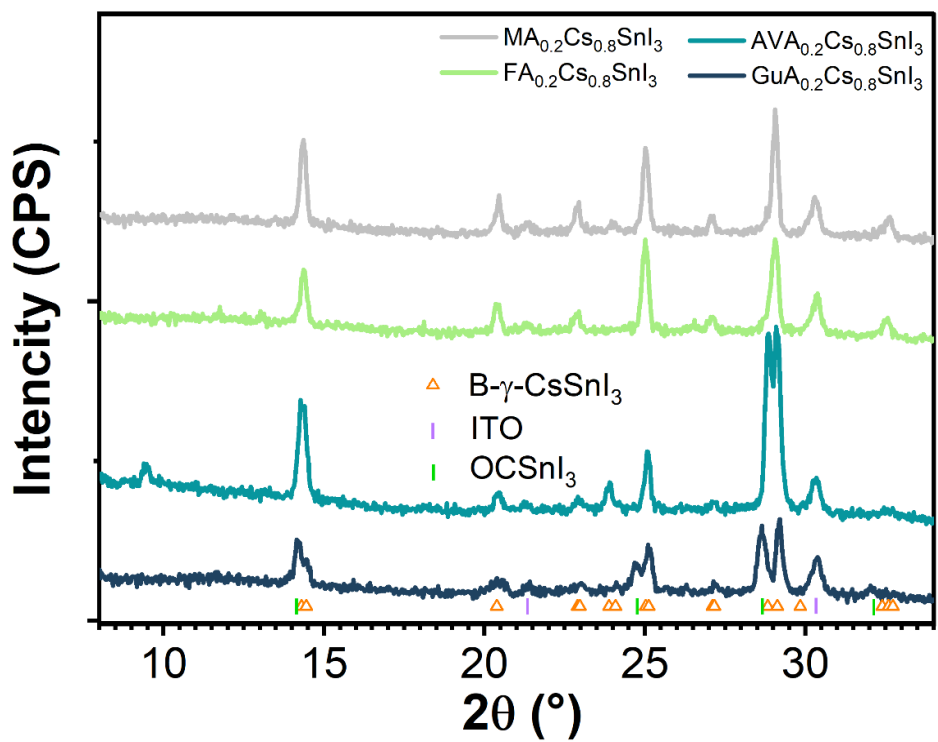

Figure 2 – X-Ray diffraction patterns of mixed cation tin halide perovskite films

The substrate peaks were identified at 21.35° and 30.38°. The primary phase in all samples corresponds to black orthorhombic *Pnma(62)* B-γ-CsSnI₃ [Entry #00-043-1162]. The incorporation of the GuA cation results in the splitting of the primary diffraction lines of B-γ-CsSnI₃, accompanied by a shift toward lower angles. This shift suggests the formation of a lattice with increased inter-planar distances, complementary to B-γ-CsSnI₃. Notably, hybrid tin perovskites, such as *P4mm(99)* α-MASnI₃ [Entry #96-433-5633] exhibit similar lattice characteristics. These findings indicate that GuA cation incorporation doesn't lead to the formation of a solid solution within the CsSnI₃ lattice but induces phase segregation, resulting in distinct GuASnI₃ and CsSnI₃ phases. Conversely, the incorporation of FA, AVA, and MA cations successfully integrates into the *Pnma* crystal lattice of B-γ-CsSnI₃. The sample containing AVA exhibits a peak at 9.48°, corresponding to structures based on the AVA spacer molecule with an inter-planar distance of 9.34 Å. In contrast, samples with FA and MA display a preferential orientation of B-γ-CsSnI₃ crystals along the (110) plane, as evidenced by the absence of diffraction peaks at 28.84°. Rietveld's method was used to calculate lattice parameters of $OC_{0.2}Cs_{0.8}SnI_3$ perovskite samples **(Tab. 2).** Unit cell volume ($V_{unit\ cell}$) of full inorganic CsSnI₃ is 929.47 Å³, organic cations expanded unit cell up to 931.18 Å³ for $FA_{0.2}Cs_{0.8}SnI_3$. The expansion of the crystal lattice upon the incorporation of large organic molecules confirms the successful integration of cations within the B-γ-CsSnI₃ lattice. However, in the case of the GuA cation, the coexistence of two perovskite phases resulted in lattice contraction of the primary B-γ-CsSnI$_3$ phase, reducing the unit cell volume to 927.6 Å$^3$.

Table 2. Rietveld calculated lattice parameters

| Sample | a (Å) | b (Å) | c (Å) | V $_{unit\ cell}$ (Å$^3$) |
|---|---|---|---|---|
| MA$_{0.2}$Cs$_{0.8}$SnI$_3$ | 8.721 | 8.672 | 12.279 | 928.64 |
| FA$_{0.2}$Cs$_{0.8}$SnI$_3$ | 8.729 | 8.697 | 12.266 | 931.18 |
| AVA$_{0.2}$Cs$_{0.8}$SnI$_3$ | 8.755 | 8.686 | 12.242 | 930.95 |
| GuA$_{0.2}$Cs$_{0.8}$SnI$_3$ | 8.689 | 8.606 | 12.407 | 927.60 |
| CsSnI$_3$ (reference) | 8.688 | 8.643 | 12.378 | 929.47 |

Sheet resistance measurements (R$_{sheet}$) by the four-probe method are crucial for evaluating charge transport properties in thin-film semiconductors, directly reflecting the film's conductivity, potential impact of defect density, and suitability for device applications. The obtained R$_{sheet}$ values (**tab.3**) revealed enhanced conductivity for samples with MA-cation (446.7 Ohm/sq), progressively reduced conductivity with larger cations FA (2021.6 Ohm/sq) and GuA (2229.5 Ohm/sq) and significantly suppressed transport properties for the sample with larger spacer AVA-cation (>100000 Ω/sq). The AVA cation promotes the formation of low conductivity 2D interlayers, limiting inter-grain charge transport in the lateral direction and drastically increasing R$_{sheet}$ [23,24].

Table 3. Resistivity of the fabricated samples

| Composition | Mean.sq. resistance (Ohm/sq) |
|---|---|
| MA$_{0.2}$Cs$_{0.8}$SnI$_3$ | 446.7 ± 158,4 |
| FA$_{0.2}$Cs$_{0.8}$SnI$_3$ | 2021.6 ± 333.1 |
| GuA$_{0.2}$Cs$_{0.8}$SnI$_3$ | 2229.5 ± 351.2 |
| AVA$_{0.2}$Cs$_{0.8}$SnI$_3$ | > 10$^5$ |

To investigate the impact of cation engineering of transport properties, we measured dark volt-ampere characteristics (JV curves) for the fabricated p-i-n diodes based on fabricated CsSnI$_3$ configurations (**fig.3**). The obtained JV data showed typical diode behavior with a strong correlation of the rectification properties on chemical composition of perovskite film. Dark JV curves have four main regions with corresponding regimes of operation: shunt current (I), recombination current (II), diffusion current (III), and contact resistance (IV)[25]. The analysis of minimal dark current densities (J$_{min}$) at 0 V exhibited the lowest value for AVA$_{0.2}$Cs$_{0.8}$SnI$_3$ (8x10$^{-8}$ A/cm$^2$), slight increase for FA$_{0.2}$Cs$_{0.8}$SnI$_3$ up to 4.0x10$^{-7}$ A/cm$^2$ and raise for MA$_{0.2}$Cs$_{0.8}$SnI$_3$ and GuA$_{0.2}$Cs$_{0.8}$SnI$_3$ to ~10$^{-5}$ A/cm$^2$. In parallel, variation of A-site composition contributed to the leakage current (J$_{leakage}$, bias= -0.1 V). FA$_{0.2}$Cs$_{0.8}$SnI$_3$ diode exhibited J$_{leakage}$ of 6x10$^{-6}$ A/cm$^2$. In contrast to small values of $J_{min}$, $MA_{0.2}Cs_{0.8}SnI_3$ and $AVA_{0.2}Cs_{0.8}SnI_3$ had increased current leakages in range of 3-4 x$10^{-5}$ A/cm². $GuA_{0.2}Cs_{0.8}SnI_3$ devices showed the weakest properties, with $J_{leakage}$=2 x$10^{-4}$ A/cm². The calculated values of shunt resistance ($R_{sh}$) represented similar trends as for leakage current. Maximum values of ~$10^4$ Ohm*cm² were obtained for $FA_{0.2}Cs_{0.8}SnI_3$ and $AVA_{0.2}Cs_{0.8}SnI_3$, the reduction in one order of magnitude was calculated for $MA_{0.2}Cs_{0.8}SnI_3$ (5x$10^3$ Ohm*cm²) and drop to 5x$10^2$ Ohm*cm² for $GuA_{0.2}Cs_{0.8}SnI_3$. We analyzed the changes of diode properties in II and III regions via fitting of a 2-diode model[26] with extraction of non-ideality factor ($m_1$, $m_2$), reverse saturation currents ($J_{01}$, $J_{02}$) and series resistance ($R_s$), as presented in **tab.4**. The corresponding calculation and fitting procedure were performed using equations **S2-S9** (**ESI**). According to the general basics of semiconductor diode physics[27–29], the dark saturation current is ruled by recombination in the devices and can be used to analyze the efficiency of the charge carrier transport. We used p-i-n diode architecture to compare the properties of various $CsSnI_3$ modifications with cation engineering. The primary recombination process of charge carriers and traps is described by non-ideality factor (m). Interpretation and analysis of the p-i-n diodes based on microcrystalline films is complex. For standard p-n architectures, the m-value is 1 for the dominating diffusion current in the space charge region. An increase in non-ideality factor up to 2 describes the domination of the recombination current. So, the reduced m-value points to the improved efficiency of the charge carrier collection and suppressed recombination dynamics. Halide perovskite-based devices have two heterojunctions with charge-transport layers and typically can't be described with a single diode circuit. For our case, we used the model with two in-series connected diodes with total m-value as a sum of each one. Devices with FA-, MA- and AVA- cations had the calculated values of non-ideality factors in the range of 1<m<2[30,31], consequently, the values rose from 1.276 for $FA_{0.2}Cs_{0.8}PbI_3$ to 1.954 for $MA_{0.2}Cs_{0.8}SnI3$. GuA modification had the largest m=2.399. The obtained trend represents the increasing impact of recombination processes with variation of the $CsSnI_3$ modification. The $FA_{0.2}Cs_{0.8}PbI_3$ configuration exhibited enhanced diode characteristics, including strong shunt resistance and low contact resistance. Relevant values of the *m*, $J_0$, and leakage current were also obtained for AVA-based devices. However, the contact resistance increased by an order of magnitude (60.8 Ohm*cm²) compared to the FA- configuration, likely due to the reduced intrinsic conductivity of the film. Despite the structural and optoelectronic similarities between FA- and MA-modified $CsSnI_3$, device analysis indicates increased recombination dynamics and degraded diode characteristics in the last one. The obtained features highlight the critical role of inter-grain boundaries in charge-transport performance[32,33]. The other principal specifics were obtained for $GuA_{0.2}Cs_{0.8}SnI_3$, where structural defects and phase segregation led to weak shunt resistance and a significant contribution of non-radiative recombination processes.

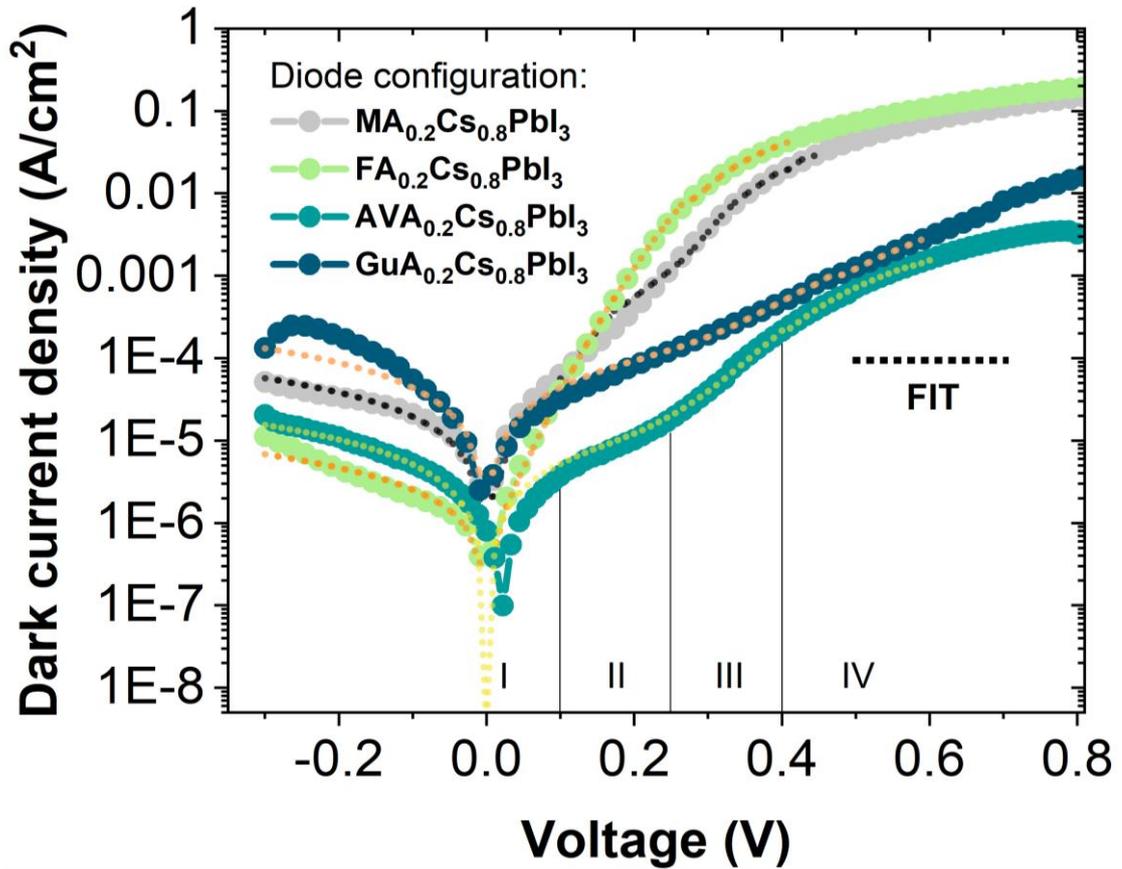

Figure 3 – Dark JV curves for p-i-n diodes fabricated with various CsSnI$_3$ modification with fitting using 2-diode model

Table 4. Calculated parameters for fitted dark JV curves for p-i-n diodes based on modified CsSnI$_3$

| Device configuration | $m_1$ | $m_2$ | $m$ ($m_1+m_2$) | $J_{01}$, A*cm$^{-2}$ | $J_{02}$, A*cm$^{-2}$ | $R_{sh}$, ohm*cm$^2$ | $R_s$, ohm*cm$^2$ |
|---|---|---|---|---|---|---|---|
| MA$_{0.2}$Cs$_{0.8}$SnI$_3$ | 0.977 | 0.977 | 1.954 | 8.2E-7 | 2.9E-4 | 5.33E+03 | 2.6 |
| FA$_{0.2}$Cs$_{0.8}$SnI$_3$ | 0.847 | 0.429 | 1.276 | 4.1E-7 | 1.8E-4 | 4.68E+04 | 1.1 |
| GuA$_{0.2}$Cs$_{0.8}$SnI$_3$ | 1.267 | 1.132 | 2.399 | 4.3E-4 | 1.2E-3 | 5.51E+02 | 12.1 |
| AVA$_{0.2}$Cs$_{0.8}$SnI$_3$ | 1.142 | 0.596 | 1.738 | 4.3E-9 | 1.2E-6 | 1.94E+04 | 60.8 |

Further, we used the unipolar structure with hole transport configuration (p-n-p) to determine the trap concentration ($N_{trap}$) and trap-filling threshold voltage ($V_{TFL}$) via analysis of JV performance in space-charge limited current regime (**fig. S6, ESI**) utilizing **eq. S10** for $N_{trap}$ evaluation. This method[34,35] is widely utilized for perovskite thin-films and reveals distinct trends with various A-site substitutions[34]. The extracted $V_{TFL}$ characterizes the voltage at which injected carriers completely fill available trap states. Variations in $V_{TFL}$—ranging from 0.10 V (MA) to 0.03 V (AVA)—likely originate from differences in band-gap values, whereas increased $N_{trap}$ (~10$^{15}$ cm$^{-3}$ for GuA and AVA) reflects higher defect densities induced by bulky cations at inter-grain interfaces.

Table 5. Calculated parameters for p-n-p diodes based on modified $CsSnI_3$

| Device configuration | $V_{TFL}$, V | $N_{trap}$, cm$^{-3}$ |
|---|---|---|
| $MA_{0.2}Cs_{0.8}SnI_3$ | 0.10 | 8.6E+14 |
| $FA_{0.2}Cs_{0.8}SnI_3$ | 0.05 | 5.7E+14 |
| $GuA_{0.2}Cs_{0.8}SnI_3$ | 0.07 | 8.7E+15 |
| $AVA_{0.2}Cs_{0.8}SnI_3$ | 0.03 | 6.7E+15 |

The analysis of recombination dynamics and impact of the ionic species in halide perovskites is complex. Charge carriers typically have fast response timescales (nano to microseconds), while ion motion introduces slow components[36–39]. To distinguish rapid electronic processes and ionic effects in p-i-n diodes based on modified $CsSnI_3$, we measured time-resolved current transients (**TRCT**, **fig.4**). A transient profile typically has a rise (as the stimulus is applied) and a fall (as the device returns to equilibrium). Analysis of TRCT curves reveals information about charge carrier dynamics, trap states, and recombination mechanisms. We performed the measurements at 100 kHz frequency and forward bias of 1V using signal generator and oscilloscope.

All device configurations showed a π-shape form of the transient profiles with switching period in the sub-microsecond range. As reported[40], $CsSnI_3$ films are characterized with very fast nonradiative recombination and carrier lifetimes on the order of a few nanoseconds. This reflects a high trap density, so any transient measurement should consider the large impact of trap-limited recombination, rather than ideal band-to-band recombination. The time corresponding to rise and fall regimes ($\tau_R$; $\tau_F$, respectively) was derived from the difference between 10 and 90% of the saturation signal values. $FA_{0.2}Cs_{0.8}SnI_3$ device showed the fastest rise response with $\tau_R$=0.28 μs, while diode with $AVA_{0.2}Cs_{0.8}SnI_3$ exhibited an increase up to 1.06 μs. Consequently, $MA_{0.2}Cs_{0.8}SnI_3$ and $GuA_{0.2}Cs_{0.8}SnI_3$ also showed reduced dynamics of the response with $\tau_R$=0.46 μs and 0.70 μs, respectively. The observed data clearly indicated a clear relation between cation modification of perovskite thin-films with rise-mode transients. In contrast, the fall profiles of TRRC revealed a slight difference in $\tau_f$. MA- and FA- based modification demonstrated very close values 0.62-0.63 μs. For $GuA_{0.2}Cs_{0.8}SnI_3$ $\tau_f$ was 0.69 μs and the slowest decay was measured for $AVA_{0.2}Cs_{0.8}SnI_3$ – 0.73 μs. Under applied forward bias, the current increases rapidly, due to the sudden narrowing of the depletion region and then continues to increase gradually if additional charge is being stored. A slow rising component in the capacitance could suggest the notable impact of the ionic motion from the charged defects (uncompensated organic cations, etc.)[41]. The measured rise profiles had rapid region with the fast increase in the signal magnitude followed by a slow exponential tail prior saturation. A fast rise indicates that most capacitive components (depletion width reduction) respond immediately, implying traps either fill very quickly or aren't significantly involved during the injection[42,43]. The period of the rapid capacitance increase in the TRCT was estimated in terms of 120 – 150 ns for the studied device structures. The recovery phase (capacitance decay) had a fast initial drop representing the free-carrier recombination. This changed by a slow tail originated from the trap states releasing carriers over a longer period. Trapped carriers don't immediately leave or recombine when the bias is removed; instead, they get thermally emitted from traps into the bands, contributing to a delayed capacitance change[44]. A close inspection of the fall-mode TRCT showed a reduced impact of the slow component for MA- and FA- modified $CsSnI_3$ devices compared to the GuA and AVA samples.

Notably, for each consequent switch of the operation regime (electrical bias on/off), we observed an overshoot (so-called spike)[45], which indicated a transient charge current opposing the expected direction due to ionic polarization[46] and/or interface recombination effects[47]. We observed the first spike with negative direction immediately after application of the forward bias. The second spike with a positive amplitude was measured at the start fall regime corresponding to the reconfiguration of the charge flow without external bias. Besides, a further spike in the decay profile was observed at the fast-to-slow charge emission transition. So, in parallel to electronic conductivity mechanism, we observed the ionic pseudo-conductivity, which produced phase-shifted currents appearing as negative capacitance. Recombination or positive-ion outflow leads to a decrease in capacitance (as charge is lost from the junction), whereas the relaxation of negative ionic charge towards the junction can cause a rise[48–50]. Microcrystalline halide perovskites typically are characterized with continuum of trap states or band tail states that cause a distributed response[51,52]. The time period of the overshoot signal took approximately 30-40 ns, so the spike signals could point to the rapid carrier's recombination via charged shallow traps, acting as capture centers. As reported [53–55], acceptor defects in $CsSnI_3$ perovskites represented tin, iodine and cesium vacancies, which could affect the overshoot transient process.

Using the TRCT data, using **eq. S11 and S12,** we calculated effective capacitances values (**fig.4(c)**), which showed comparable charge storage abilities (~$3.0 \times 10^{-8}$ F). Exceptionally, AVA substitution demonstrated significantly lower capacitance (~$2 \times 10^{-9}$ F), reflecting reduced stored charge possibly due to suppressed diffusion capacitance linked to poor inter-grain transport. Reverse switching loss measurements (**fig.4(d), eq.S13-S14**, 80–800 µW/cm² at 100 kHz) identified AVA as superior (~100 µW/cm²), offering minimal energy dissipation beneficial for high-frequency applications[56], whereas the MA composition exhibited the highest loss (~800 µW/cm²), indicating substantial energy inefficiency during switching.

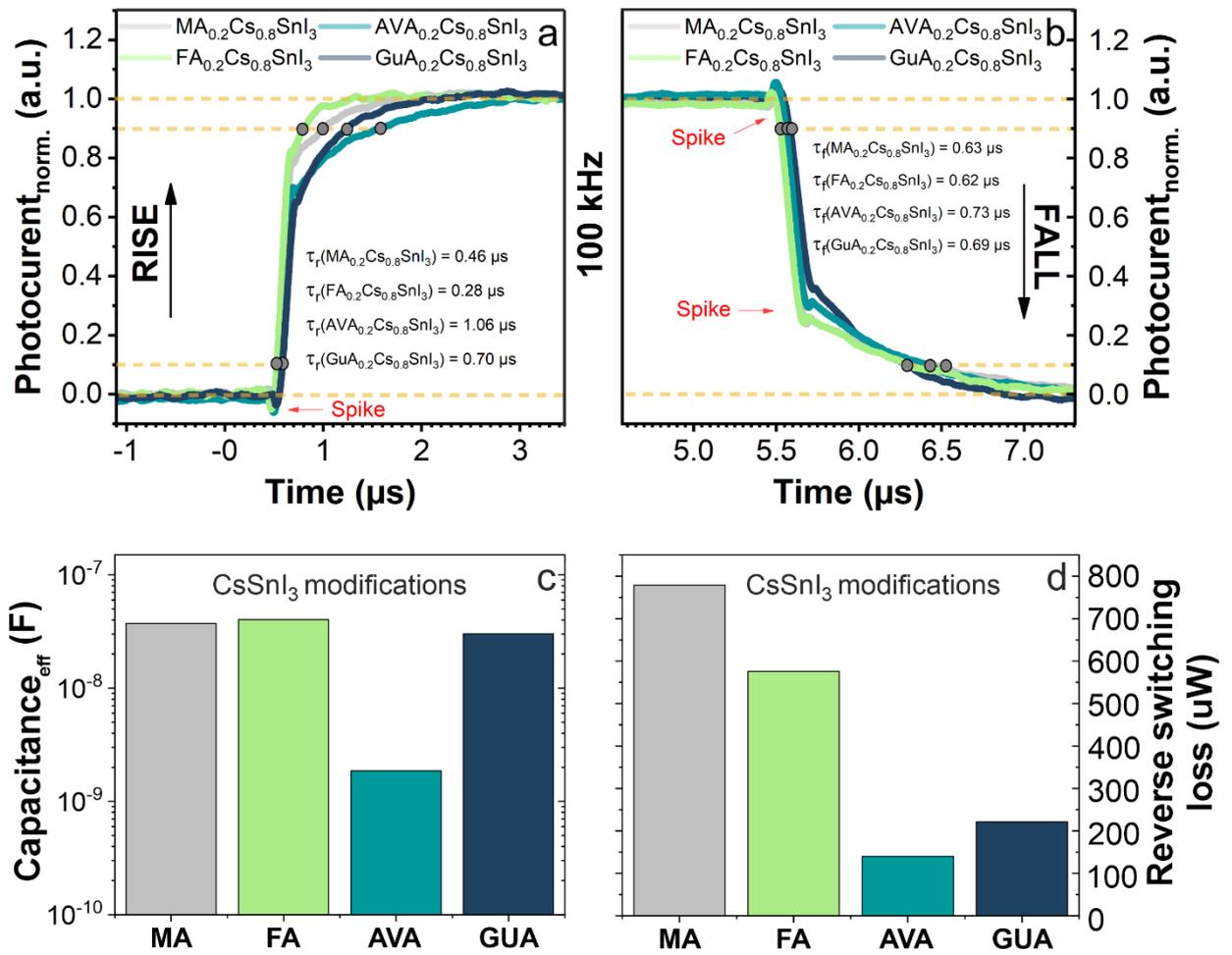

Figure 4 – The transient-current measurements for p-i-n diodes in rise mode (a), fall mode (b), histograms for calculated $C_{eff}$ values (c), histograms for calculated reverse switching losses (d)

In conclusion, A-site cation engineering in $CsSnI_3$ thin films was confirmed as an effective approach to precisely control structural and electronic properties, significantly influencing conductivity, defect states, diode characteristics, and transient responses. Although MA-substitution yielded the highest conductivity ($R_{sheet}$ ~450 Ohm/sq), elevated defect densities (~$8.6 \times 10^{14}$ $cm^{-3}$), increased recombination losses, and reduced shunt resistance limited efficiency of the charge transport in the devices. FA-substitution emerged as an optimal balance, offering moderate conductivity, minimized trap concentrations, superior diode characteristics and rapid transient responses. Conversely, bulky GuA and AVA cations introduced substantial structural modifications, including phase segregation (GuA) and formation of low-conducting 2D interlayers (AVA), drastically suppressing lateral charge transport or defect accumulation at the interfaces. Notably, AVA-based devices showed the lowest effective capacitance (~$2 \times 10^{-9}$ F) and minimal reverse switching losses (~100 $\mu W/cm^2$), advantageous for high-frequency and energy-efficient applications despite slower transient responses. The observed findings emphasize that precise A-site cation selection can strategically optimize CsSnI$_3$ film performance tailored to specific device requirements.

## SUPPLEMENTARY MATERIAL

The supplementary material includes description of the materials section (materials, inks preparation, device fabrication, laser patterning cycle, characterization), AFM images and KPFM maps of perovskite surface. The data that supports the findings of this study are available within the article and its supplementary material.

The authors gratefully acknowledge the financial support from the Russian Science Foundation with project № 22-79-10326.

## Conflict of Interest

The authors have no conflicts to disclose.

## Author Contributions

**D.S.S. and P.A.G.** conceived the work.

**G.V.S, P.A.G, A.A.Z.** performed experiments on perovskite devices and the electro characterization.

**A.P.M.** performed measurements of the dynamic response.

**L.L.O.** performed microscopy measurements.

**D.S.S. and P.A.G.** provided administrative support and resources.

**D.S.S.** coordinated the research activity.

The manuscript was written with contributions from all the authors. All the authors approved the final version of the manuscript.

The Electronic Supplementary Information (ESI) for the paper:

Modulation of Charge Transport and Rectification Behavior in CsSnI$_3$ Thin Films Through A-site Cation Engineering

Anna A. Zarudnyaya[1], Gleb V. Segal[1], Lev O. Luchnikov[1], Andrey. Luchnikov[1], Andrey P. Morozov[1], Sergey Yu. Yurchuk[2], Pavel A. Gostishchev[1*] and Danila S. Saranin[1*]

[1]LASE – Laboratory of Advanced Solar Energy, NUST MISiS, 119049 Moscow, Russia

[2]Department of semiconductor electronics and device physics, NUST MISiS, 119049 Moscow, Russia

**Corresponding author:** Dr. Pavel A. Gostishchev gostischev.pa@misis.ru and Dr. Danila S. Saranin saranin.ds@misis.ru

**Experimental section:**

*Materials*

All organic solvents – dimethylformamide (DMF), dimethyl sulfoxide (DMSO), isopropyl alcohol (IPA), chlorobenzene (CB) were purchased in anhydrous, ultra-pure grade from Sigma Aldrich, and used as received. 2-Methoxyethanol was purchased from Acros Organics (99.5+%, for analysis), HNO$_3$ (70%). Devices were fabricated on In$_2$O$_3$: SnO$_2$ (ITO) coated glass (R$_{sheet}$<7 Ohm/sq) from Zhuhai Kaivo company (China). NiCl$_2$·6H$_2$O (from ReaktivTorg 99+% purity) used for HTM fabrication. Tin iodide (SnI2, >99.9%), Cesium iodide (99.99%), trace metals basis from LLC Lanhit, Russia and Methylammonium iodide (MAI, >99.99% purity from GreatcellSolar), Formamidinium iodide (FAI, >99.99% purity from GreatcellSolar), 5-Ammonium valeric acid iodide (AVAI, >99.99% purity from GreatcellSolar), Guanidinium iodide (GuAI, >99.99% purity from GreatcellSolar), were used for perovskite ink. Fullerene-C60 (C60, >99.5%+) was purchased from MST NANO (Russia). Bathocuproine (BCP, >99.8% sublimed grade) and Poly-TPD were purchased from Osilla Inc. (UK).

*Inks preparation*

A nickel oxide precursor ink was prepared by dissolving NiCl$_2$·6H$_2$O powder in 2-methoxyethanol (2-ME) at a concentration of 35 mg/mL. To 1 mL of this solution, 20 μL of 70% nitric acid (HNO$_3$) was added. The poly-TPD solution was prepared by dissolving poly-TPD in chlorobenzene (CB) to achieve a concentration of 4 mg/mL. The solution was stirred at 50°C for 1 hour. A 1.3 M perovskite precursor solution, comprising organic cations (OC) such as methylammonium (MA), formamidinium (FA), 5-ammoniumvaleric acid (5-AVAI), and guanidinium (GuA), was prepared by mixing OCI:CsI:SnI$_2$ in a 0.2:0.8:1 ratio. All perovskite precursor solutions were dissolved in a solvent mixture of dimethylformamide (DMF) and dimethyl sulfoxide (DMSO) in a 4:1 ratio. The solutions were stirred for 30 minutes and subsequently filtered before deposition. For the hole-blocking layer, a BCP solution was prepared by dissolving BCP in isopropyl alcohol (IPA) at a concentration of 0.5 mg/mL and stirring overnight at 50°C.

*Sample preparation*

*Thin-film samples*

Thin-film samples for absorption, four-probe and profilometry measurements were fabricated on a glass (Soda Lime 1.1 mm) with a structure glass/perovskite. For XRD, KPFM and AFM measurements, ITO coated glass substrates (1.1 mm) were used with ITO/perovskite structure of the samples.

*Devices*

**p-i-n** diodes for dark JV and space charge limited current measurements were fabricated with inverted planar architecture ITO coated glass/c-NiO /perovskite/C60/BCP/Bi-Cu.

Unipolar **p-n-p** devices for time-resolved current transient TRCT measurements were fabricated with architecture ITO coated glass/c-NiO /perovskite/Poly-TPD/BCP/Bi-Cu.

*Layers deposition routes*

Firstly, the ITO substrates were cleaned with acetone and IPA in an ultrasonic bath and activated under UV-ozone irradiation for 30 min. The NiO HTL was spin-coated in an air atmosphere with a relative humidity not exceeding 40% with the following ramp: (1 s – 500 rpm, 10 s – 4000 rpm). The deposited NiO layer was annealed at 125 °C for 15 minutes and at 310 °C for 1 hour. Perovskite films were crystallized on top of HTL with a solvent-engineering method. The deposition and crystallization processes of perovskite layers were conducted inside glove box with an inert nitrogen atmosphere. Perovskite precursors were spin-coated with the following ramp: step 1: 10 sec, 1000 rpm. Step 2: 30 sec, 4000 rpm. Antisolvent CB with volume of 200 µl was dropped on substrate at 25 sec of the second step. Then, substrates were annealed at 50 °C (20 min) and 100 °C (20 min) for conversation into the black perovskite phase. As an ETL 40 nm of C60 was deposited with the thermal evaporation method at $10^{-6}$ Torr vacuum level. The Poly-TPD layer was spin-coated at 4000 rpm (30 s) and annealed at 100 °C (5 min) inside a glove box. The BCP layer was spin-coated at 4000 rpm (30 s) and annealed at 50 °C (1 min) inside a glove box. Metal cathode Bi (15 nm) and Cu (85 nm) thick was also deposited with thermal evaporation through a shadow mask to form a 0.15 cm² active area for each pixel.

*Characterization*

Surface roughness and films thicknesses were measured with KLA-Telencor stylus profilometer. X-Ray diffraction **(XRD)** of perovskite layers was investigated with diffractometer Tongda TDM-10 using CuKα as a source with wavelength 1.5409 Å under 30 kV voltage and a current of 20 mA. absorbance spectra **(ABS)** of perovskite films were carried out via SE2030-010-DUVN spectrophotometer with a wavelength range of 200–1100 nm. Atomic force microscopy **(AFM)** and Kelvin probe force microscopy **(KPFM)** measurements were performed in room ambient conditions using Ntegra (NT-MDT) microscope. NSG10/Pt (Tipsnano) probes were used with tip curvature radius 30 nm. KPFM study utilized an amplitude modulation regime. AFM and KPFM images were obtained simultaneously by two-pass method.

Dark JV curves and space charge limited current were measured in an ambient atmosphere with Keithley 2400 SMU in 4-wire mode and a settling time of $10^{-2}$ s. The time-resolved current transients (TRCT) was measured with DIGILENT Analog Discovery Pro 3450 (2 units), which were used as oscilloscope and pulse generator.

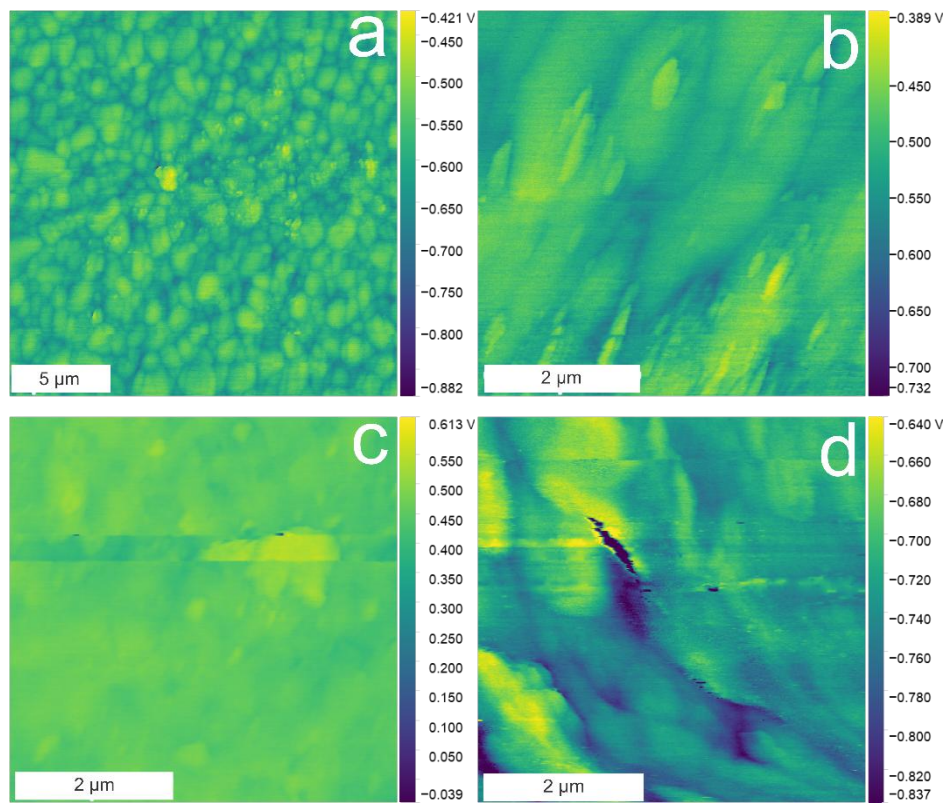

Figure S1 – maps of surface potential For OC$_{0.2}$Cs$_{0.8}$SnI$_3$ perovskite with MA (a), FA(b), AVA(c), GuA(d) cations

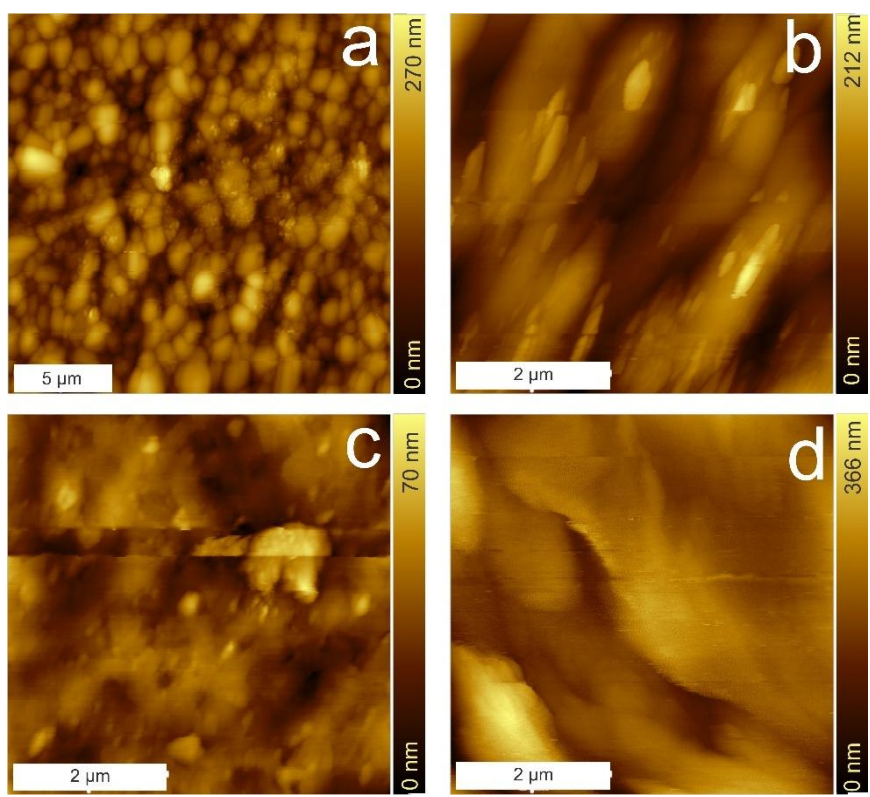

Figure S2 – surface topography of OC$_{0.2}$Cs$_{0.8}$SnI$_3$ perovskite with MA (a), FA(b), AVA(c), GuA(d) cations

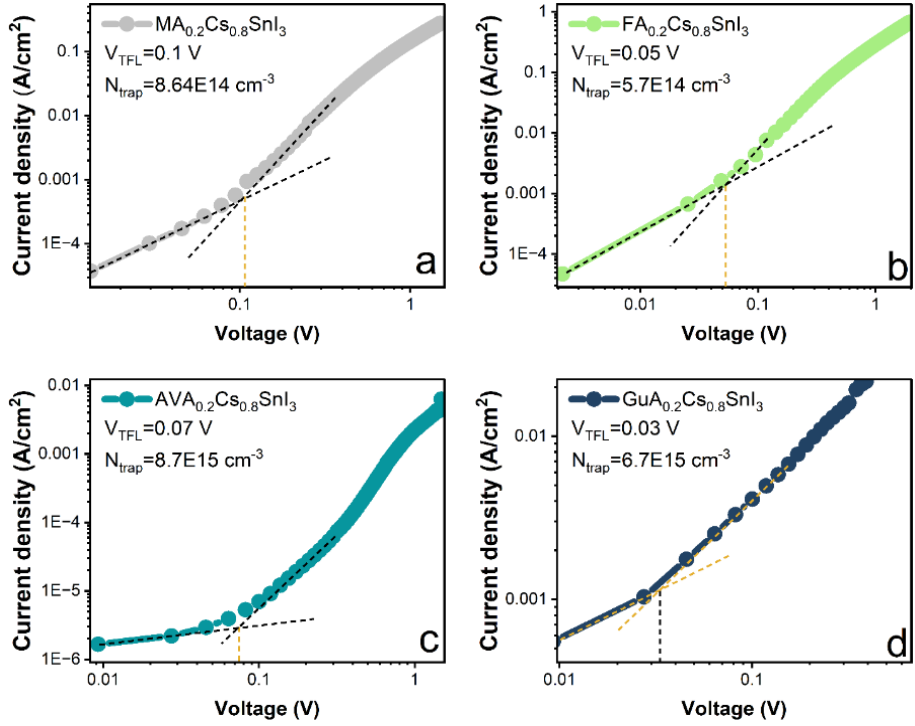

Figure S3 – JV curves for p-n-p diodes based on modified $CsSnI_3$

**Measurement and Calculation Methodology**

*Tauc plots analysis*

The Tauc plots calculation was realized using equation:
$$(\alpha h\nu)^2 = A(h\nu - E_g) \qquad (S1)$$
where α is absorption coefficient being a function of wavelength $\alpha(\lambda)$, $h$ is Planck constant, $E_g$ is an optical band gap of a semiconductor, $\nu$ is frequency, $A$ is proportionality constant, and $n$ is Tauc exponent.

*Four-probe measurements*

We measured the series resistance using four-point probe station "VIK-UES" (MEDNM, Russia). Four in-line tungsten carbide probes with a spacing of 0.75 mm are used, the applied current was 100 uA. Each sample was measured in the center four times, to obtain an average series resistance of the film and estimate the spread. Systematic error in the measurement system does not exceed 2%.

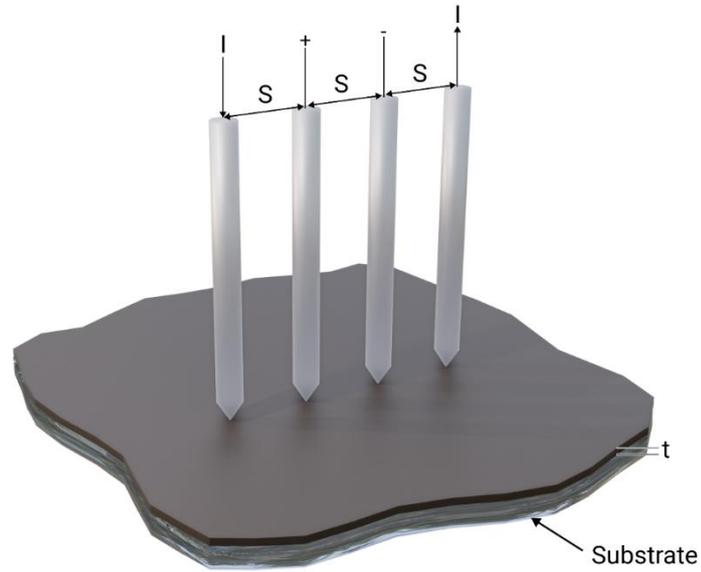

Figure S4 – scheme of four-probe measurement

### 2- diode model fitting

The equation for the diode current-voltage curve, including series (Rs) and shunt (Rsh) resistance, has the following expression (S1):

$$J = J_0 \cdot \left( \exp\left( \frac{q(V - J \cdot R_s)}{m \cdot k \cdot T} \right) - 1 \right) + \frac{V - J \cdot R_s}{R_{sh}} . \quad (S2)$$

For common pn junction, such a model describes the characteristics of real structures quite accurately. However, for a p-i-n structure, such a model does not always allow us to express the characteristic corresponding to experimental results. The reason for this is the presence of two barriers from the p- and n-regions, therefore, to calculate the current-voltage characteristics of the p-i-n structure, we used a two-diode model, represented by the equivalent circuit in Fig. S2. [2]

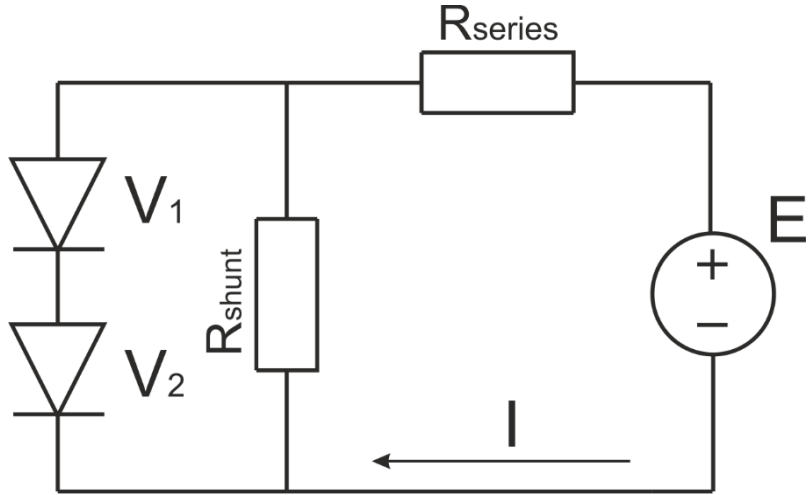

**Figure S5** – Double diode circuit for modeling of pin PSC

Diodes 1 and 2 in this circuit are ideal diodes, the current-voltage characteristics of which are described by the expressions S2-S3:

$$J_1 = J_{01} \cdot \left( \exp\left( \frac{qV_1}{m_1 \cdot k \cdot T} \right) - 1 \right) \tag{S3}$$

$$J_2 = J_{02} \cdot \left( \exp\left( \frac{qV_2}{m_2 \cdot k \cdot T} \right) - 1 \right) \tag{S4}$$

Two diode structures are connected in series, so the currents are equal to each other.

$$J_d = J_{01} \cdot \left( \exp\left( \frac{qV_1}{m_1 \cdot k \cdot T} \right) - 1 \right) = J_{02} \cdot \left( \exp\left( \frac{qV_2}{m_2 \cdot k \cdot T} \right) - 1 \right) \tag{S5}$$

Since the equivalent circuit is branched, it is necessary to solve a system of equations obtained from Kirchhoff's laws to calculate the current-voltage characteristic, For a given voltage V, unknown values are the currents in the diode and shunt resistance circuits $J_d$ and Rsh. The voltages applied to each diode $V_1$ and V2, and the total current J, which we must find for each given voltage. The system of equations (S4) - (S7) is sufficient for the numerical calculation of the current – voltage characteristics according to the two-diode model, but it is necessary to calculate the model parameters: leakage currents of each individual diode $J_{01}$ and $J_{02}$, non-ideality coefficients $m_1$ and $m_2$ of each diode, series resistance $Rs$ and shunt resistance $Rsh$.

$$V_1 + V_2 = V - J \cdot R_s \tag{S6}$$

$$J_{R_{sh}} = \frac{V - J \cdot R_s}{R_{sh}} \tag{S7}$$

$$J = J_d + J_{R_{sh}}. \tag{S8}$$

The calculation was done with one of the methods for multi-parameter optimization, in which the objective function requiring minimization of the sum for the differences between the experimental and theoretically calculated currents (S8).

$$\sum_{i=1}^{m}(J_{ex_i} - J_{teor_i})^2, \tag{S9}$$

where m is the number of experimental points.

Calculation program was developed in Borland Delphi 7, which allows determining the parameters of a two-diode structure using multi-parameter optimization by the coordinate descent method.

*Space charge limited current measurements*

Determination of the trap-filling voltage ($V_{TFL}$):

Find the threshold voltage $V_{TFL}$ (see Fig. S6) at which the slope of the log(J)-log(V) plot increases significantly from region I to region II.

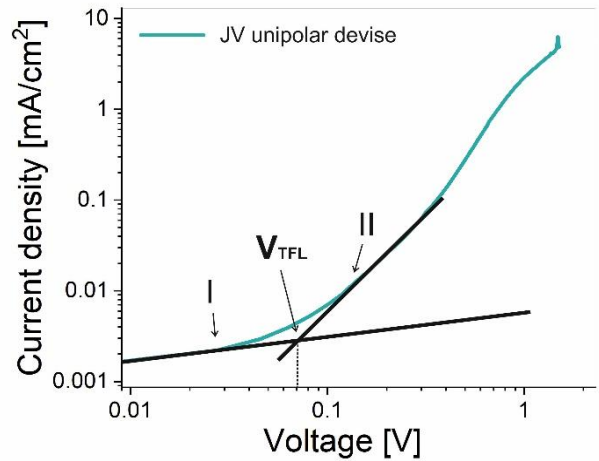

Figure S6 – Determination of $V_{TFL}$ from the unipolar device JV plot

Calculation of trap density ($N_t$):

$$N_t = \frac{2\varepsilon\varepsilon_0 V_{TFL}}{eL^2} \quad (S10)$$

where $\varepsilon$ is relative dielectric permittivity of perovskite, $\varepsilon_0$ – vacuum dielectric constant (8.854·10$^{-12}$ F·m$^{-1}$), $e$ – elementary charge (1.602·10$^{-19}$ C), L – perovskite layer thickness, $V_{TFL}$ – trap filling voltage

### *Effective Capacitance*

The effective capacitance of a diode represents the incremental change in stored charge with respect to a change in applied voltage. This parameter encompasses two key contributions. First, the depletion (junction) capacitance (Cj) arises from the charge separation in the depletion region; under reverse bias, as the depletion width increases, the capacitance decreases, with its value largely determined by the doping concentrations and geometric structure of the junction. Second, the diffusion capacitance (Cd) occurs under forward bias when minority carrier injection results in a significant buildup of excess charge in the quasi-neutral regions. Since this charge storage is governed by dynamic processes—carrier injection, diffusion, and recombination—the diffusion capacitance is typically much larger than the depletion capacitance at moderate to high forward bias. We defined effective capacitance as:

$$C_{eff} = \frac{dQ}{dV}, \quad (S11)$$

where Q is the total charge (sum of the depletion and diffusion charges).

In our study, rather than measuring capacitance directly, we calculated $C_{eff}$ indirectly from the diode's recovery behavior. Using the relationship for the RC time constant,

$$t = RC_{eff}, \quad (S12)$$

Where t is the reverse recovery time.

R is the resistance of the device measured at its operating forward bias.

We derived effective capacitance from the measured reverse recovery time and the resistance present during switching. This approach enables us to capture the dynamic response of fast-switching diodes while minimizing the influence of parasitic elements. The calculated effective capacitance thus reflects how rapidly the diode can charge or discharge during switching events—a critical factor that impacts both switching speed and associated power losses.

For our subsequent calculations—particularly in estimating the effective capacitance from the measured RC time constant and calculating the reverse switching loss—we focus on the reverse recovery time. This is because the reverse recovery interval is when the diode expends energy to remove the stored charge, and the corresponding RC time constant $t_{rr}=R \cdot C_{eff}$ reflects the dynamic behavior under switching conditions. Importantly, the resistance used in this RC calculation is the forward dynamic (differential) resistance, measured from the slope of the forward JV curve, since it characterizes the conduction path through which the stored charge is removed. Although the JV curve in the reverse region shows a much higher resistance (owing to the widened depletion region), that high resistance pertains to the diode's blocking state rather than the transient process of charge removal.

### *Reverse Switching Loss*

The reverse switching loss quantifies the energy dissipated when a diode transitions from its conducting state to a non-conducting state. During this transient, the diode must remove the stored minority carrier charge accumulated under forward bias. In our experiments with perovskite diodes—where the diode is switched from a forward-biased condition (approximately 1 V) to zero voltage—the switching loss per cycle is approximated by the area under the reverse recovery transient.

Assuming a triangular approximation for the recovery current waveform, the energy loss per cycle $E_{rr}$ is calculated as:

$$E_{rr} = \frac{1}{2} I_{rr} V_R t_{rr} \tag{S13}$$

$I_{rr}$ is the peak reverse recovery current (measured at the point when the diode voltage is 1 V),

$t_{rr}$ is the reverse recovery time (the time taken for the diode to remove the stored charge during turn-off), and

$V_R$ is the voltage swing during the transition.

In our case, since the diode is switched from a forward bias of 1 V to zero volts, we take $V_R=1$.

To determine the average power loss due to reverse recovery, we multiply the energy loss per cycle by the switching frequency f:

$$P_{rr} = E_{rr} f \tag{S14}$$

For our measurements, the diode is driven at a switching frequency of 100 kHz. The transient response is captured using a high-speed oscilloscope and appropriate current probes so that $I_{rr}$ and $t_{rr}$ can be accurately extracted from the waveform.